# Spectroscopy, Manipulation and Trapping of Neutral Atoms, Molecules, and Other Particles using Optical Nanofibers: A Review

Michael J. Morrissey[1], Kieran Deasy[2], Mary Frawley[2,3], Ravi Kumar[2,3], Eugen Prel[2,3], Laura Russell[2,3], Viet Giang Truong[2] and Síle Nic Chormaic[1,2,3,*]

1. School of Chemistry and Physics, University of Kwa-Zulu Natal, Durban 4001, South Africa; E-Mail: morrissey@ukzn.ac.za (M.M.)
2. Light-Matter Interactions Unit, OIST Graduate University, 1919-1 Tancha, Onna-son, Okinawa 904-0495 Japan; E-Mails: kieran.deasy@oist.jp (K.D.); mary.frawley@oist.jp (M.F.); ravi.kumar@oist.jp (R.K.); eugen.prel@oist.jp (E.P.); laura.russell@oist.jp (L.R.); v.g.truong@oist.jp (V.G.T); sile.nicchormaic@oist.jp (S.N.C)
3. Physics Department, University College Cork, Cork, Ireland

* Author to whom correspondence should be addressed; E-Mail: sile.nicchormaic@oist.jp (S.N.C.); Tel.: +81-98-966-1551.

**Abstract:** The use of tapered optical fibers, i.e., optical nanofibers, for spectroscopy and the detection of small numbers of particles, such as neutral atoms or molecules, has been gaining ground in recent years. In this review, we briefly introduce the optical nanofiber, its fabrication and optical mode propagation within. We discuss recent progress on the integration of optical nanofibers into laser-cooled atom and vapor systems, paying particular attention to spectroscopy, cold atom cloud characterization and optical trapping schemes. Next, a natural extension on this work to molecules will be introduced. Finally, we consider several alternatives to optical nanofibers that display some advantages for particular applications.

## 1. Introduction

Quantum mechanics plays a crucial role in the development and understanding of future technologies governed by quantum rules. The isolation of single atoms allows researchers to directly observe quantum properties of light-matter interactions. In recent years researchers have been investigating the possibility of detecting, controlling and manipulating quantum systems, such as cold atoms [1], trapped ions [2, 3] and molecules [4] close to the surface of nanostructured devices. These devices includes microcavities [5], atom chips [6], superconducting circuits [4], and optical nanofibers [1]. The interest in these devices primarily arises from their potential to offer methods by which single atoms/particles can be trapped, probed and manipulated, thereby providing a useful tool for advancing quantum engineered devices.

The focus of this review paper is on the application of optical nanofibers (ONF) [7-10] in the detection, manipulation and trapping of laser-cooled atoms and the role they plays in this and related research fields. Optical fiber sensors are a well-established tool [11] and, more recently, the versatility of ONFs as sensing tools [12, 13] for very small numbers of particles is also becoming evident. We begin in this section by presenting the basic concept of the ONF and its fundamental properties. Section 2 introduces the ONF as a probe for cold atoms as well as atoms in a vapor. In these cases the nanofiber can act as either an active (absorption measurements) or passive (fluorescence measurements) probe to detect the atoms, and can be used to investigate a large number of interesting physics phenomena. Also in this section, the various techniques proposed to precisely trap and manipulate atoms are presented, as well as the atom-surface effects under such conditions. Aside from the progress in cold atom-fiber interaction experiments, ONFs have also found applications in sensing molecules, quantum dots, and nanodiamonds – areas that are reviewed in Section 3. In Section 4 some alternative systems for single atoms and molecules based on micro- or nanostructuring within the ONF itself are presented. The paper concludes with some comments on the future direction of this research field.

*1.1 Properties of an optical nanofiber*

An ONF is a circular dielectric waveguide whose diameter is smaller than the wavelength of the light which propagates within it. Such an ONF can be realized as the waist region of a biconical tapered optical fiber consisting of three distinct regions: (i) the normal fiber, (ii) the transition or taper region, and (iii) the waist, as illustrated in Fig. 1. To fabricate such a device the same general fabrication technique is typically applied, involving the heating of an optical fiber to a molten state while elongating it to create a tapered optical fiber. Until now a number of different heat sources have been used for this process, including a gas flame [14, 15], $CO_2$ laser [16], microheater [17], and electrical strip heater [18]. Independent of the type of heat source a pair of parameters are vital for determining the shape of the tapered region, along with the length and diameter of the ONF waist region [19, 20]. These parameters are: (i) the size of the hot zone and (ii) the elongation length of the fiber. For this reason the flame brushing technique is usually considered to be the preferred technique for incorporating optical nanofibers into cold atoms systems [14] since both of these parameters can be controlled with relative ease.

**Figure 1.** Schematic illustration of a sample optical nanofiber with a 125 µm outer diameter and a core diameter of 4.4 µm. The narrowest region has a diameter in the range of hundreds of nanometers. Typical taper lengths are in the region of 6 cm.

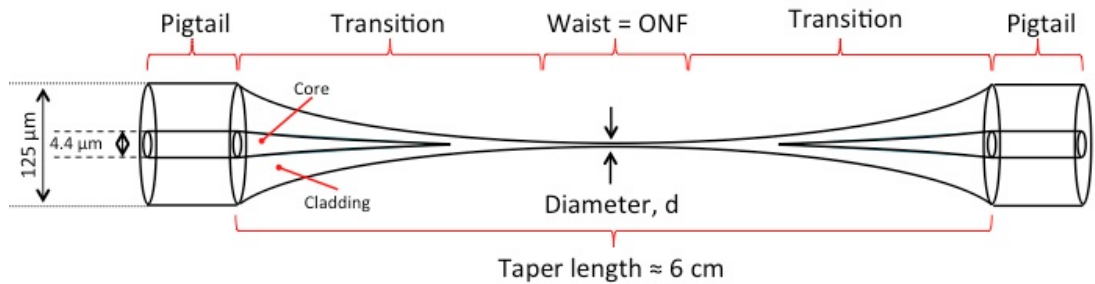

Another important factor in the fabrication of high performance ONFs is that optical losses due to the tapered regions should be kept to a minimum. For the achievement of this goal the adiabaticity criteria should be met [21]. These criteria take into account the changing profile of the fiber within the transition region and determine the best profile available. Assuming the diameter of the core and cladding decrease at the same rate, the core is reduced in size to the point where it has negligible influence on the guided modes within the ONF [22]. Thus, the guiding properties of the ONF are determined by the difference in the refractive indices between the original cladding, $n_{clad}$, and the ambient environment, $n_{amb}$.

In many applications of ONFs it is necessary for the fiber to operate in single mode, although interest in higher order mode propagation is increasing due to some advantages offered in modal interference for particle trapping [23-25]. The condition for a fiber to be single mode is given by the V-number, which depends on the fiber diameter as well as the numerical aperture of the fiber, such that:

$$V = \frac{k_0 d \sqrt{\left(n_{clad}^2 - n_{amb}^2\right)}}{2} < 2.405 \ , \qquad (1)$$

where $k_0 = 2\pi/\lambda$ is the free space wave-number and $d$ is the fiber diameter. This single mode condition is illustrated in Fig. 2, where only the fundamental, $HE_{11}$, mode exists for values of the V-number lower than 2.405. This condition is met experimentally by ensuring the ONF diameter is below the single-mode cut off diameter.

Due to the sub-wavelength diameter of the ONF, the evanescent field [26] extends further into the optically rarer medium than in a conventional optical fiber. The light intensity on the surface can easily exceed the intensity inside the ONF if its diameter is small enough. Moreover the polarization dependence of the decaying behavior in the evanescent field is more distinct in this case, and the field intensity varies azimuthally around the fiber depending on the polarization of the propagating light. The enhancement of the evanescent field lays a favorable platform for the interaction with matter and also permits ONFs to be used for ultra-sensitive sensing applications.

## 2. Neutral Atoms

The potential for using cold atomic ensembles in the evolution of quantum technologies is undisputed. However, it is clear that further advances in the precision control and manipulation of cold atoms are essential. Since the development of laser cooling and trapping of neutral atoms in the 1980's [27-29], many techniques have been successfully developed to trap [6, 30-33], manipulate [34-40], and probe [41-44] cold atoms. Both the trapping and manipulating of atoms can be performed using either magnetic fields or far-off resonance laser light, while the probing is most conveniently performed with near- or on-resonant light. In recent years, ONFs have attracted considerable interest in the field of quantum optics due to their ability to efficiently coupling light and matter, thus having the ability to simultaneously trap, manipulate and probe neutral atoms [45]. There has also been a proposal on incorporating an ONF into an optical lattice to create small, cold atom samples with control over the final atom number [46].

**Figure 2.** A plot of the effective index of refraction, $n_{eff} = \beta/k_0$, against V-number for an optical fiber with cladding and core refractive indices as 1.4537 and 1.000, respectively. $\beta$ represents the propagation constant. The vertical line indicates the boundary between single- and multimode guidance in the fiber (V = 2.405).

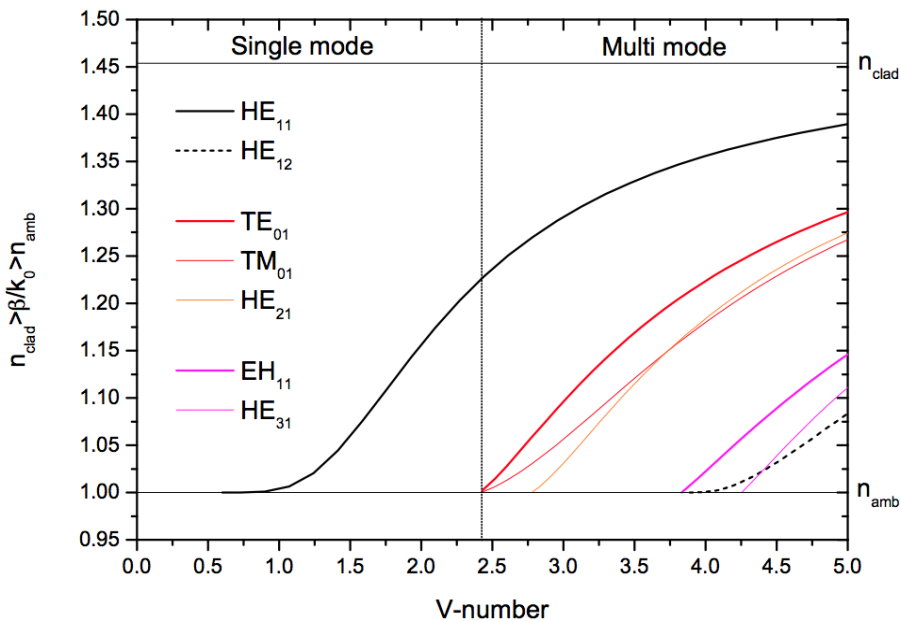

Spontaneous emission from cold, neutral atoms that are sufficiently close to a sub-wavelength diameter fiber can couple into the guided modes of the fiber [47, 48]. Due to multiple surface effects, which are described in Section 2.1, this spontaneous emission can be strongly enhanced and the resultant fluorescence can be utilized to passively probe atoms in the vicinity of the ONF surface. This effect is utilized by two fluorescence techniques, namely emission and excitation fluorescence detection, which are described in Sections 2.2 and 2.3 respectively. The ONF can also be used as a detection device in an active manner whereby near- or on-resonant light propagating via the ONF's evanescent field is absorbed by the neutral atoms. These three cold atom probing techniques are illustrated in Fig. 3. The absorption detection method (Fig. 3(c)) can also be extended to detect atoms in a vapor cell as is described in Section 2.5. By taking advantage of the evanescent field properties of the ONF, the combination of far-off-red and blue-detuned evanescent fields can be used to create optical potentials, which can be used to trap and manipulate cold atoms. Variations of such techniques are described in Section 2.6.

**Figure 3.** Three techniques to detect neutral atoms using ONFs. **(a)** Emission fluorescence detection, **(b)** Excitation fluorescence detection [49], **(c)** Absorption detection [50].

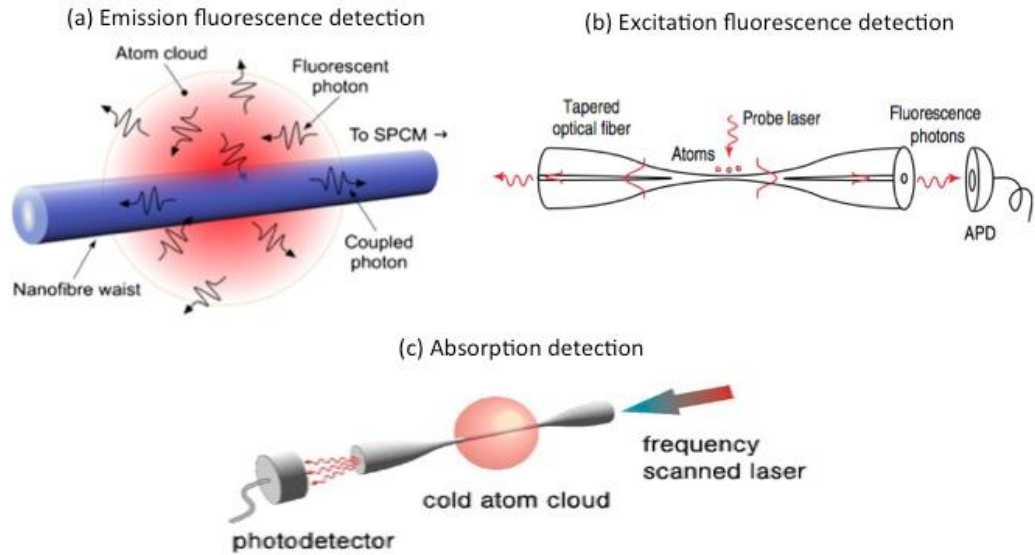

*2.1 Cold atom-surface interaction*

The study of individual neutral atoms in the vicinity of dielectric surfaces has gained renewed interest in recent year [47, 49, 51, 52]. Regarding ONFs, this is primarily due to their ability to manipulate atoms near their surfaces making them ideal for the development of devices in quantum optics and nanotechnology. In order to develop these technologies it is imperative to have a thorough understanding of how the presence of the ONF dielectric surface affects the behavior of the atom.

It is well known that the presence of any material body can substantially modify the spontaneous decay rate of an atom [47, 53-56]. Spontaneously emitted photons close to the surface of the ONF can be radiated into space or they can be coupled into the guided modes of the ONF [57]. The presence of the ONF itself enhances the coupling into the guided modes. The enhancement of the decay rates is largest when the atom is on the fiber surface and the effect reduces as the atoms are positioned further from the ONF surface. The decay rates not only depend on the position of the atom relative to the fiber surface but also on the fiber size itself. The enhancement becomes significant when the diameter of the ONF is small in comparison with the radiation wavelength [47]. The enhancement also slightly varies for different magnetic sublevels. In a realistic system a coupling efficiency of up to 28% of the spontaneously emitted photons can be achieved for a Cs atom near the surface of an ONF [47]. Such a high efficiency offers a promising technique for single atom detection.

The dominant electromagnetic interactions that take place in this context are the van der Waals [58] and Casimir-Polder forces [59]. The van der Waals force for ONFs can be viewed as an attractive force that pulls atoms towards the surface of the fiber and operates at distances $<\lambda/10$ from the fiber surface, where $\lambda$ is the wavelength of the radiation The Casmir-Polder force is also an attractive force, but only becomes significant at distances $>\lambda/10$ from the surface. In the context of cold atoms, these forces have been measured for several surfaces [60, 61] and can be exploited to create trapping potentials as discussed in Section 2.6.

The aforementioned atom-surface interactions cause a shift or perturbation in the spectral emission of excited atoms close to the surface of the ONF. Russell *et al.* [59] theoretically studied this effect by modeling the spectral properties of atoms whose spontaneous emission couples to the guided mode of the ONF. Here the fiber was treated as a planer surface. It was determined that for typical ONF diameters (200-600 nm), the fluorescence excitation spectrum exhibits a well-pronounced asymmetry with red-side broadening as well as red-detuned shifting the of peak position caused by the van der Waals effect. The inclusion of the Casimir–Polder effect has minimal influence on the asymmetry of the line shape, but slightly reduces the red-shift of the peak position. In addition it was determined that the asymmetry becomes more pronounced for atomic ensembles that are tightly confined around the optical nanofiber. Later, Frawley *et al.* [62] considered the effect of the fiber curvature on the van der Waals interaction with an atom. The predicted red-side-broadening effect has been experimentally observed by Sagué *et al.* [50] using absorption detection (see Section 2.4 for more details of the results). Here, the asymmetries in the absorption profile of atoms interacting with the evanescent field of an ONF have been attributed to the van der Waals frequency shift.

The red-shifted profile was also observed by Nayak *et al.* in 2008 [49], when the emission spectrum of atoms was coupled into the guided mode of a fiber and a long red tail in the emission spectrum was observed (see Section 2.3 for further details on the measurement technique) which can be seen represented by Trace B in Fig. 4. This was initially attributed to the presence of atoms close to the surface and seemed to agree with theoretical predictions [63, 64]. However, when Nayak *et al.* [65] performed a more systematic investigation of atoms in a surface-bound potential, this effect was found to be due to atoms on the surface itself rather than free, cold atoms. The authors first measured the laser induced fluorescence spectrum of cold atoms and investigated how this evolved over time for changing fiber surface conditions. The line shape of the excitation spectrum changes from Lorentzian for free, cold atoms (Trace A in Fig. 4) and over time the spectrum broadens to a large red tail line shape (Trace B in Fig. 4). This implies that, as time progresses, more and more atoms form bound-states with the fiber and induce a spectral line broadening effect until the narrow peak near the atomic resonance is due to the contribution from free-atoms with a broad spectrum in the red-detuned side due to the bound-atoms. The effect plays a crucial role in determining the surface conditions and leads to a reduction to the amount of emission that is coupled to the guided modes of the fiber (as can be seen in Fig. 4 when the relative amplitudes of the signals are compared). In fact, it is suggested that the main source of the red tail is produced by room-temperature atoms falling into the surface potential rather than the cold atoms themselves. Thus, the background density determines the speed at which atoms fall into the surface-potential. The atoms can be removed from the bound potential with the aid of a violet laser and this can be used to control the surface conditions. In this manner the observed spectrum returns to its original state and the normal Lorentzian spectral emission spectrum can be once again achieved. This provides a method by which the effect of surface conditions can be eliminated.

2.2 *Emission fluorescence detection of cold atoms*

The detection and quantitative analysis of cold atomic ensembles are essential analytical tools required in the development of quantum technologies. The simplest method for cold atom detection is fluorescence imaging, where the atoms are irradiated with resonant laser light and the resultant emissions of the atoms are focused on either a CCD camera or a photodiode. Using a combination of these devices, parameters such as size, number of atoms and density profile can be easily determined with sufficient accuracy for most experiments. Dynamic properties of the experimental setup such as loading rate, lifetime and decay rate can also be determined.

**Figure 4.** Fluorescence excitation spectra from cesium measured through an optical nanofiber for the closed-cycle transition, 6S $F = 4 \leftrightarrow$ 6P $F = 5$. Traces A and B correspond to without and with the effect of violet laser irradiation, respectively. [49]

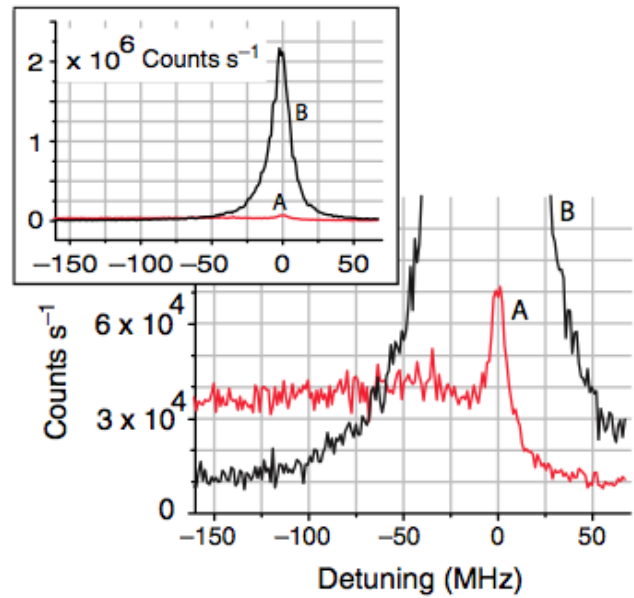

For normal operational parameters of a magneto-optical trap (MOT), atoms in the cloud absorb light from the cooling laser beams stimulating the atoms to an excited state. For atoms that are close to the surface of the ONF, a portion of the resultant spontaneous emission can couple directly into the guided mode of the fiber waist [66, 67]. With the aid of a single photon counter module (SPCM) attached to the end of the fiber, photons can be detected and the specific attributes of the cold atom system can be determined. The high coupling efficiency of fluorescence photons to the guided modes enables fluorescence measurements from a very small number of atoms. This is done while the atoms are continuously trapped in the MOT and thus a continuous real-time measurement is observed. This fluorescence detection scheme of cold atoms is illustrated in Fig. 3(a).

The detection technique was first implemented by Nayak *et al.* [68] whereby the coupling of the fluorescence from Cs atoms into the ONF was observed simply by monitoring the signal on the SPCM as the laser beams and magnetic field are sequentially switched on and off. A large increase was observed when the MOT was switched on due to the spontaneous emissions coupling to the guided modes. From experimental parameters of the setup, it was determined that the effective number of atoms contributing to the signal at any given time was five. This accurately agreed with the photon count amplitude observed due to the presence of cold atoms around the fiber, indicating the potential of such devices for detection of a low number of atoms. Again, using this technique, the density profile, size and shape of the atomic cloud was determined by magnetically translating the cold atom cloud across the waist of the ONF and measuring the photon count rate as a function of cloud position. All the fiber fluorescence imaging results were compared to those achieved using standard fluorescence imaging techniques using photodiodes and CCD cameras and found to be in good agreement.

This emission fluorescence detection method was adopted by Morrissey *et al.* [69] where the technique was used to determine the dynamic loading rates and life-times. This was done by monitoring the coupled photons as a function of time as the MOT loaded from the background and decayed due to internal atom collisions when the atom source was switched off. The results were compared to those taken simultaneously using standard MOT measurement techniques and were found to be in good agreement. Using this same fiber fluorescence technique Russell *et al.* [70, 71] measured temperatures above and below the Doppler limit using two different methods, one being that of forced oscillations [70] and the other being that of release-recapture [71]. Observed temperature variation corresponds with the limits of normal MOT behavior. Again, good agreement was found between temperature measurements made using the optical nanofiber and conventional fluorescence imaging methods with a photodiode.

*2.3 Excitation fluorescence detection of cold atoms*

Nayak *et al.* [68] developed an alternative technique to measure the fluorescence of cold atoms with the aid of an ONF. Once the cold atoms are prepared in the MOT, the trapping beams are switched off and, for a short period of time, the cold atoms are excited by a resonance probe beam. During this excitation period the atoms absorb from the probe beam resulting in spontaneous emission of photons, some of which couple to the guided mode of the fiber and are counted by an avalanche photodiode (APD). The probe beam is then switched off and the MOT beams are switched on allowing the atomic cloud to reload from the background. This experimental setup is illustrated in Fig. 3(b). In this way the observation is accumulated over many cycles. In this excitation fluorescence experiment the estimated atom number in the observation region is reduced by a factor of 70 when compared to direct emission fluorescence detection. This is due to the expansion of the atom cloud while the cooling beams are off during the excitation period. However, this method has the advantage of being insensitive to scattering from the irradiating light.

Using this technique Nayak *et al.* [49] were able to measure the excitation fluorescence spectrum and detected surface effects which have already been referred to in Section 2.1 [65, 68]. This method can be extended to the detection of single atoms around the ONF [49]. In this case, the atom number is dramatically reduced by varying experimental parameters of the MOT. The excitation fluorescence that couples into the guided modes is split at the fiber output using a 50/50 nonpolarizing beam splitter and detected using two separate SPCMs. The photon correlation between the two channels is measured by performing a Hanbury-Brown and Twiss experiment. The photon coincidences clearly display anti-bunching effects confirming the detection of single atoms using the ONF. Under these conditions the excitation spectrum was measured to further understand the atom behavior. For a low intensity of the probe beam the spectrum exhibits almost a Lorentzian shape but is slightly asymmetric with a small red tail with no power broadening observed. This indicates that the spectrum is induced by free atoms in the vicinity of the ONF. However, there exists a small dip at the central peak. This dip is more pronounced for larger probe intensities and an additional broadening effect is observed. This dip was attributed to the mechanical effect due to scattering from the probe beam. It is worth noting that single atom behavior was only observed for clean fibers (after irradiation of violet laser) – a further indication of the crucial role of surface interactions in such experiments.

*2.4 Absorption detection of cold atoms*

Absorption imaging is a standard detection technique in cold atom physics whereby properties of the cold atom ensemble can be determined by measuring the absorption of a resonant light field. The strong evanescent field surrounding the waist of a sub-wavelength tapered fiber allows light-matter interactions with media surrounding the fiber and therefore facilitating in-fiber spectroscopy on the cold atom ensemble which surrounds the ONF. Given that the evanescent field decays exponentially from the surface, typically within a distance of $\lambda/2\pi$, this method can also be used to investigate surface interactions.

Work published in 2006 by Kien *et al.* [72] highlights the need for knowledge regarding the optical response of an atom in a resonant field which propagates along the surface of a fiber. The authors demonstrate that, for low field intensity of the evanescent field, when the atom is in the close vicinity of the fiber surface, the scattered power can be up to 60% of the propagation power. When the fiber diameter is comparable to the wavelength of propagation, the light is mainly scattered into free-space while scattering into the guided modes is weak. In the case of high propagation powers, a further increase in the propagation power will lead to a dramatic decrease in the scattering efficiency due to the saturation effect. This is a positive outcome as, even with small input powers, circulation intensities of the probe beam at the ONF waist are very high. This implies that a wide range of non-linear effects are obtainable with minimal probe powers.

The first experimental demonstration of in-fiber spectroscopy of a cold atomic sample was published in 2007 by Sagué *et al.* [50]. Atoms are first captured and cooled in a standard MOT while the probe laser is switched off. The MOT cooling and repump laser beams, as well as the magnetic field, are switched off and the probe laser is switched on. During this time period the probe beam frequency is scanned around the atomic cooling transition allowing the atoms to absorb the light propagating in the evanescent field and the signal is measured on an avalanche photodiode (APD). This measurement technique is illustrated in Fig. 3(c). A series of absorption spectra are shown in Fig. 5 for three evanescent field intensities. The presence of a mode propagating through the fiber increases spontaneous emission by approximately 57% at the surface - an effect that had not been observed before to such an extent without a cavity. For probe powers larger than 100 pW, the line shapes are narrower than expected. This is explained by the effect of the light-induced dipole forces on the density of the atomic cloud. For distances smaller than 370 nm, i.e. in the region that contains more than 75% of the evanescent field power, the largest integrated density of the atomic cloud is predicted in the case of zero detuning. For blue (+3 MHz) and red (-3 MHz) detunings, this integrated density is lowered due to the effect of the light-induced dipole forces. This results in reduced absorbance and leads to an effective line narrowing. The measured linewidths approach 6.2 MHz for vanishing powers. This result exceeds the natural Cs D2 linewidth in free space by almost 20%. This broadening can be explained by surface interactions detailed in Section 2.1, i.e. the van der Waals shift of the Cs D2 line and the modification of the spontaneous emission rate of the atoms near the fiber. Both effects have the same magnitude and only their combination yields the very good agreement between the theoretical model and the experimental data. Other surface effects such as the red-detuned-shifting of the center of the absorption profile could not be measure due to the drift of the probe laser frequency. They conclude by highlighting that the subwavelength diameter fiber can be used to detect, spectroscopically investigate, and mechanically manipulate extremely small samples of cold atoms. On resonance, as little as two atoms on average coupled to the evanescent field surrounding the fiber and absorbed 20% of the total power transmitted through the fiber.

**Figure 5.** Line shapes obtained from an evanescent field around a nanofiber embedded in a laser-cooled sample of Cs. The measured linewidths approach the natural linewidth (6.2 MHz) for vanishing probe powers. [50]

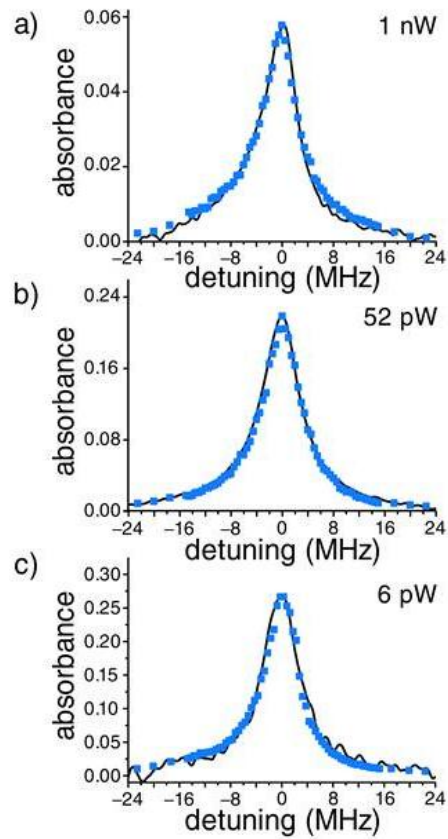

Thus, by interacting with few numbers of atoms via an appreciably high-powered evanescent field, single and few atoms in the evanescent region can behave as an optically dense system. This is the subject of the work in Hakuta *et al.* [73]. They explore the atom-field interaction around a nanofiber using laser-cooled, Cs atoms and find that the atom/nanofiber interaction may open a new technique to trap single atoms without any external field. By using this trapping technique, they experimentally investigate small number of atoms by observing the fluorescence excitation spectrum.

A publication in 2008 by Nayak *et al.* [49] highlighted that, since an appreciable amount of propagating radiation is distributed in the evanescent region, single atoms in this region work as a good nonlinear medium due to their optical density. The photoabsorption spectrum is measured for small number of atoms, revealing a possibility to realize an optically dense system using hundreds of atoms. For example, photoabsorption through the nanofiber reaches about 50% when one atom sits on the surface - if several atoms are prepared on the nanofiber surface, the system may become optically opaque. It should be noted that, when the atom is positioned away from the surface the absorption becomes smaller obviously and one may need more atoms to realize the optically dense medium.

*2.5 Absorption detection of vapor gas*

The concept of in-fiber spectroscopy for cold atoms can easily be extended to include atoms in a vapor. In this case, transit-time broadening results in significantly broadened line shapes as a hot atom will pass through the evanescent field in under 1 nanosecond, compared to a cold atom which has a transit time on the order of microseconds [74]. In 2008, Spillane *et al.* [75] discussed the observation of nonlinear interactions of a Rb vapor with an ONF-generated evanescent field using very low levels of input light. In fact, they were able to saturate the vapor using an input power level of 8 nW (Fig. 6(a)) and observe electromagnetically induced transparency (EIT) (Fig. 6(b)).

**Figure 6.** **(a)** Transmission spectrum for a nanofiber in a vapor cell for increasing probe powers (2, 4, 8, and 10 nW) for the Rb $D_2$ transition; **(b)** Saturated absorption spectrum with cross-polarized pump and probe beams for a Rb vapor cell (upper, blue) and a Rb nanofiber system (middle, green and lower, red) for the $D_1$ manifold of $^{85}$Rb. The two side peaks correspond to the Doppler-free F=3 to F'=2 and F=3 to F'=3 hyperfine transitions. [75]

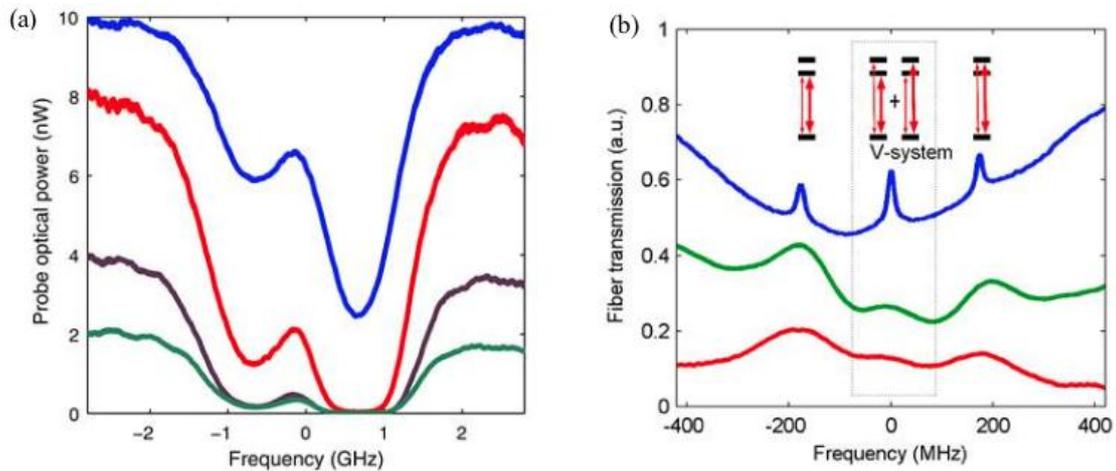

In 2009, Hendrickson *et al.* [76] observed transmission through a tapered fiber which was a nonlinear function of the incident power. This effect can also allow a strong control beam to change the transmission of a weak probe beam. This once again indicates that the ONF-atom interface can be used for nonlinear effects, EIT, slow light etc. For example, in 2010, Hendrickson *et al.*, published work about their observation of two-photon absorption in the ONF-vapor system using input light at levels below 150 nW [77]. The transit-time broadening resulting from the fast atoms passing the nanoscale waveguide produces two-photon absorption spectra with sharp peaks that are very different from conventional line shapes (see Fig. 7). Russell *et al.* [78] have since proposed using the ONF in a cloud of cold atoms for the demonstration of 1- and 2-photon absorption.

**Figure 7**.  Resonant two-photon absorption in a tapered optical fiber. The percent transmission of the 776 nm signal through the tapered fiber is plotted as a function of its detuning from the upper atomic state. (a) 780 nm power level of 146 nW. (b) 780 nm power level of 726 nW. [77]

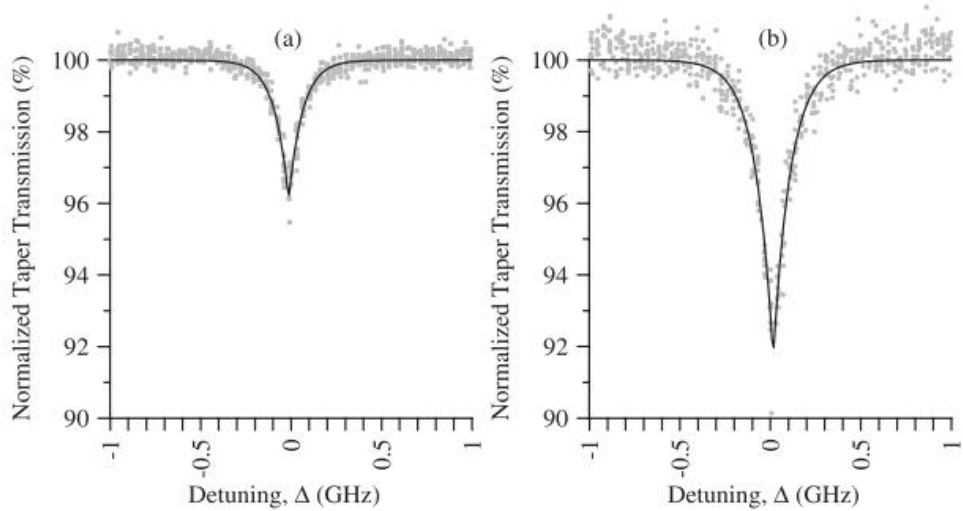

*2.6 Trapping of Neutral Atoms with ONFs*

As mentioned previously, the extension of the evanescent field from the ONF surface is very small and thus creates a high radial intensity gradient. This leads to a large gradient force on the atoms around the ONF in the transverse direction which can be utilized to create a dipole trapping potential. For red-detuned light the optical force becomes attractive and the blue-detuned light creates a repulsive force. By combining these forces along with the van der Waal's force, dipole trapping potential for atoms around the ONF can be created. To date, several trapping schemes and configurations have been developed to trap, guide and probe neutral atoms using ONFs making it a powerful tool in this research area.

The first proposal regarding the trapping of neutral atoms around an ONF was written by Balykin *et al.* [79]. In this paper it was proposed to use a red-detuned evanescent field propagating through a subwavelength diameter ONF to create an optical potential attracting atoms towards the surface. For ONFs with a diameter which are two times smaller than the wavelength of propagating light, this attractive force can be counter-balanced by the centrifugal force of atoms moving in a circular motion around the TOF thereby trapping the atoms close to the fiber surface.

Shortly after this theoretical publication, Kien *et al.* [80] proposed that neutral atoms can be trapped close to the surface of the ONF using two-color evanescent light fields. This proposed technique utilized a red-detuned light field as well as the van der Waals force, to attract atoms towards the nanofiber while a blue-detuned light field repels the atoms from the fiber. Due to the fact that the decay length of the evanescent field is wavelength dependent, by choosing the respective powers of the red- and blue- detuned beams a radial potential close to the nanofiber surface is created. If one or both fields are linearly polarized, it results in the formation of two local minima azimuthally around the fiber. This configuration allows atoms to be confined in two lines parallel to the fiber axis. If the input light fields are circularly polarized, a ring shaped potential is formed around the fiber resulting in atoms being confined to a cylindrical shell around the fiber.

Sagué *et al.* [25] proposed three separate single-color techniques to trap atoms based on two-mode interference of a blue-detuned evanescent field of an ONF. The difference in phase velocities of two modes simultaneously copropagating through the fiber allows for the creation of a stationary evanescent interference pattern long the length of the fiber at a specific distance from the surface. Controlling the power distribution between the modes enables the modification of the evanescent field of each mode. This allows the creation of field minima where the two fields cancel due to destructive interference. The atoms are thus radially trapped due to the varying decay length of the modes, axially trapped due to the different phase velocities of the modes and azimuthally trapped due to the polarization matching of the modes. Combining the $HE_{11}$ and $TE_{01}$ modes, or the $HE_{11}$ and $HE_{21}$ modes, results in two periodic arrays of traps on either side of the fiber in the axial direction. The site separation in the axial direction is determined by the beat length of the modes and the radial modes are axially offset from each other by an amount inversely proportional to the beat length. The $HE_{21}$ and $TE_{01}$ mode combination creates four axial arrays of traps, which are elongated due to the larger difference in the beat lengths. The atoms are trapped about 100 – 200 nm from the surface of the ONF. The parameters of the trap were such as to achieve trap depths of 1 mK, a trap lifetime of 100 s, with an initial kinetic energy corresponding to 100 µK. Depending on the mode combination the traps can be setup using powers of 25 – 50 mW.

In 2010, Vetsch *et al.* [1] experimentally realized the two-color evanescent field to create a one dimensional optical lattice to trap Cs atoms around an ONF, work that was furthered by Dawkins *et al.* [81]. Two far-red-detuned lasers were counterpropagated through the ONF, producing an evanescent standing wave which creates an attractive force towards the fiber surface. This attractive force is balance by a single far-blue-detuned laser field. By choosing the correct power ratio between the red- and blue-detuned light fields a minimum potential is achieved at a distance of a few 100 nm from the surface of the fiber, yielding trapping frequencies of 200, 313, 400 kHz in the radial, axial and azimuthal directions respectively. The atoms are confined in trapping sites along the axis of the fiber by the standing wave and radially confined by the potential created by the red-and blue-detuned laser fields. All the laser fields were linearly polarized, thereby confining the atoms azimuthally. This experimental setup is depicted in Fig. 8(a), which includes an illustration of the lattice. Figure 8(b) shows a fluorescence image of the trapped atoms. The atoms were confined in a one dimensional optical lattice approximately 200 nm from the surface of the ONF with each lattice site separated by 500 nm. The average occupancy was 0.5, limiting the trap to about 2000 atoms per millimeter. When spectral properties of the atoms were investigated it was found that their linewidth was slightly larger than the atomic linewidth due to atom-surface interactions. This trapping technique can be easily adapted to create other configurations. Schneeweiss *et al.* [82] manually tuned the relative phase between the counterpropagating beams of the standing waves with the aid of acousto-optic modulators to demonstrate the optical transport of cold atoms along the nanofiber. Reitz *et al.* [83] proposed a double helix potential for the cold atoms using the same beam configuration as [1], only with circularly polarized beams. The helical confinement arises from the beam intensity variations in the azimuthal direction [84]. Such a configuration would be extremely difficult to achieve using free space optics.

**Figure 8.** (a) Experimental setup of the atom trap. The blue-detuned running wave in combination with the red-detuned standing wave creates a trapping potential. A resonant laser is used for probing the atoms via the evanescent field. (b) Fluorescence image of the trapped atoms. [1]

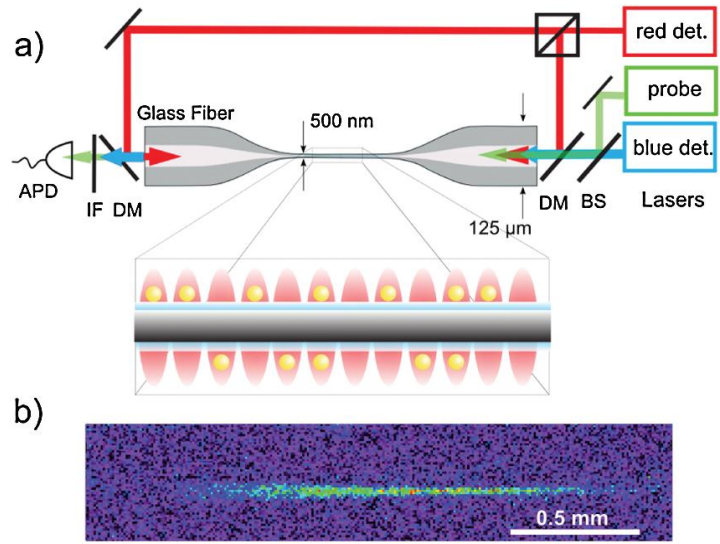

In 2012 Goban *et al.* [85] realized a two-color optical trap, which utilizes a so-called 'magic compensation' that traps the $6S_{1/2}$ ground and $6P_{3/2}$ excited states of Cs. Similar to traps mentioned above, this configuration is created by two counterpropagating beams, which are red-detuned to the magic wavelength, $\lambda_{red}$ = 937 nm, creating a standing wave along the length of the ONF. This attractive force is balance by a second pair of counterpropagating beams blue-detuned to the magic wavelength, $\lambda_{blue}$ = 686 nm. Both beams are co-linearly polarized resulting in a 3-D state insensitive optical trap with approximately 224 atoms at a distance of 215 nm from the fiber surface. The power required for the red- and blue-detuned beams are 0.4 and 5 mW respectively, creating trapping frequencies of 199, 273 and 35 kHz in the radial, axial, and azimuthal axis respectively. The trapped atoms were analyzed using the absorption techniques mentioned in Section 2.3 where a probe beam propagates through the fiber, interacts with the trapped atoms in the ONF section of the fiber, and is measured at the other end by an APD. Due to the operation of the magic wavelength compensated trap, no appreciable shift in the transition frequency (0 ± 0.5 MHz) or broadening of the linewidth (0.5 ±0.1 MHz) was observed. When this is compared to the noncompensated trap a considerable shift in the transition frequency (13 MHz) and broadening of the linewidth (14.8 MHz broader than the natural linewidth) was observed.

## 3. Molecules and Other Particles

Considering the aforementioned successes in using optical nanofibers for atom sensing, it is not surprising that they have also been used in experiments with molecules and other particles. This section gives a description of such advances, beginning with the use of nanofibers to examine spectroscopic properties of surface adsorbed particles. This offers a novel method for probing molecular photo-response, spectroscopic changes due to molecule-surface interactions, the dynamics of surface agglomeration and thin-film formation. The section then develops to encompass exciting advances in controlled microparticle, quantum dot and nanodiamond deposition onto nanofiber surfaces. Such particles can then be site-addressed to obtain a myriad of spectroscopic and behavioral data through the fiber. Furthermore, quantum dot, and particularly nanodiamond, coupling to nanofibers have promising applications in the creation of fiber output single photon sources.

*3.1 Molecular spectroscopy*

Nanofiber-based surface absorption spectroscopy of molecules was first examined by Warken *et al.* [86], who showed that the sensitivity of such a system is orders of magnitude higher than previous free-space techniques. Due to the their spectral sensitivity to molecular surface arrangement, 3,4,9,10-perylene-tetracarboxylic dianhydride molecules (PTCDA) were selected for experiment. These crystals were heated below a nanofiber, and sublimated molecules were adsorbed onto its surface. Absorption spectra were obtained through the fiber for various deposition times (0.5 to 2.3 x $10^7$ molecules), showing clear vibronic progression. Additionally, molecular condensation and agglomeration over time resulted in transmissional line shifts, allowing the authors to investigate the post-deposition film evolution on the nanofiber surface. The noted molecular agglomeration subsequently received further study [87], where the dynamics of the system under ambient and ultrahigh vacuum (UHV) conditions were compared, and significant reduction of molecular mobility in UHV was observed.

A similar deposition method was later used to excite and detect PTCDA fluorescence spectra through a nanofiber [88]. Interestingly, the absolute peak positions were shifted relative to those obtained from solution spectra. The authors associate this with the interaction of the molecules with the fiber surface, and also highlight the non-negligible contribution of self-absorption of fluorescence photons to recorded spectra. This should be minimized, or compensated for, to retrieve the expected mirror symmetry between absorption and fluorescence spectra. Non-linear experiments have also been conducted with nanofibers, demonstrating two-photon excited fluorescence measurements of adsorbed Rhodamine 6G (Rh6G) molecules. For this experiment, a solvent-dripping technique was developed to extend the applications of surface adsorption spectroscopy to a larger variety of molecules [87]. The aforementioned surface adsorption techniques are highly applicable for modeling systems of organic thin-film growth and - as they are entirely fiber-based - may be used for remote spectroscopic studies. Crucially, they facilitate recording of both absorption and fluorescence spectra for a given molecular surface coverage, and pave the way for nanofiber self-absorption free fluorescence spectroscopy on individual surface-adsorbed molecules.

Optical nanofibers have also been used in saturation absorption spectroscopy studies of acetylene ($^{12}C_2H_2$) molecules in a chamber at 200 Pa pressure [89]. Passing an infrared (IR) pump beam through the nanofiber and retroreflecting it as a probe beam, the spectra of the P9 transition were obtained with a narrow saturated signal. The dependence of the saturation parameter on nanofiber diameter was also calculated and optimized. Due to the fact that lowering the gas pressure can reduce the pressure broadened width, longer ONF lengths are desirable for such studies. Wiedemann *et al.* [90] recently presented a novel application for nanofibers where organic photochromic molecules show a reversible light-induced change of their absorption spectra and are thus ideal candidates for optical switching studies using nanofibers. Low concentrations of SpiroOH molecules, which photoswitch between transparent and colored when exposed to UV and white light, were drip-coated onto the nanofibers

State-switching was then induced by simultaneously coupling a UV LED through the fiber. Absorbance spectra confirmed the switching mechanism; this effect can be seen as a time progression in Figs. 9(a) and (b). Photobeaching, photodestruction and cyclability of the molecules were also investigated using the optical nanofibre and the authors indicate that switching speeds (0.025-1Hz) could be increased by orders of magnitude by propagating tailored laser pulses through the fiber.

**Figure 9.** Absorbance spectra of spiroOH on a nanofiber, with white light and additional UV-light exposure. All molecules are initially in the transparent state; **(a)** UV illumination increases the absorbance up to a stable point in the photostationary state; **(b)** The absorbance decreases after the UV exposure has stopped. [90]

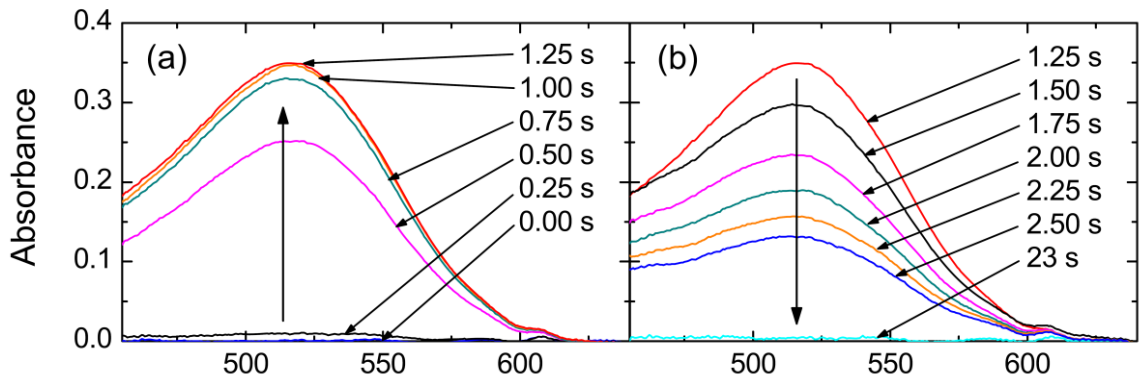

As with neutral atoms, bringing single particles into contact with the evanescent field of a nanofiber has obvious potential for both sensing and spectroscopic studies. Such a technique was demonstrated by Gregor *et al.* [91], who introduced individual charged particles to the surface of a nanofiber via a segmented linear Paul trap. The individual fluorescent dye-doped polystyrene beads were excited with the aid of a microscope objective, and preselected before being brought to the fiber surface with a success probability close to 100%. The authors compared the fluorescence spectra achieved using the microscope objective to that obtained through the fiber and calculated a seven-fold overall increase in detected fluorescence using the nanofiber. They also used FDTD simulations to relate the dip observed in fiber transmission following particle introduction to the system to the number of particles in an incident cluster. This non-contact electrospray injection method can facilitate the study of any type of charged particles which can be brought into suspension.

*3.2 Quantum Dots and nanodiamonds*

Optical nanofibers were first used for single quantum dot photoluminescence spectroscopy by Srinivasan *et al.* [92]. A single layer of InAs quantum dots (QD) was embedded in an $In_{0.15}Ga_{0.85}As$ quantum well, grown at the center of a GaAs waveguide. To increase accessibility to the QDs and reduce the overall QD number, microdisk cavities of diameter $d = 2$ μm were fabricated from the material and cryogenically cooled to 14 K. The authors compared the QD emission spectra obtained from objective lens pumping to those obtained by pumping and collecting via nanofiber coupling to the disk. This showed a 25 fold increase in the collected power through the fiber without coupling to whispering gallery modes in the disk. Spatial mapping and single QD photon collection was demonstrated by scanning the nanofiber over the surface of the disk. Focusing on photon collection efficiency, Davanco et al. [93] subsequently modeled optical nanofiber coupling to single emitters embedded in thin dielectric membranes. FDTD simulations were run for the fiber and membrane as a composite system with associated supermodes and the modified spontaneous emission rate of an embedded quantum dot was modeled as a two-level atom. The fiber collection efficiencies calculated for both vertically and horizontally oriented dipoles amount to up to 30%, which exceed those obtainable with a high NA objective by an order of magnitude To further increase the efficiency of such a system, they modeled the optical nanofiber coupling to a quantum dot embedded in a suspended semiconductor channel waveguide [94], predicting that up to 70% of the dipole's emission can be collected by the nanofiber in this system. The authors subsequently theoretically investigated nanofiber coupling to individual emitters bound to the surface of thin dielectric membranes via polymer, sol-gel or crystalline hosts [95], again yielding high collection efficiencies.

Simplifying the aforementioned systems to quantum dots attached directly to nanofibers, Garcia-Fernandez et al. [87] dripped CdSe QDs dissolved in heptane onto a nanofiber waist. Adsorbed dots were calculated to number $3 \times 10^5$, and absorption and fluorescence spectra were recorded through the fiber. In a similar theme to above, an observed shift in the fluorescence maximum was attributed to surface interactions with the nanofiber.

Yalla et al. [96] subsequently progressed this line of inquiry to observe fluorescence and emission properties of single quantum dots. To achieve this, they deposited CdSeTe (ZnS) QDs at 20 μm intervals along a nanofiber using a sub-picoliter needle dispenser, with a positioning accuracy of 5 μm. Deposition sites were individually excited using an inverted microscope, as illustrated in Fig. 10(a). Photon correlations were measured from one fiber pigtail, clearly showing anti-bunching behavior, and QD blinking was also observed. Single-step and double-step blinking corresponded to one and two QDs on site respectively where the probability of depositing a single QD using this method was estimated at 60% and fluorescence photon coupling rates into fiber guided modes were calculated. The saturation behavior of the QDs relative to excitation intensity was also addressed (see Fig. 10 (b)). The channeling of fluorescence photons into the nanofiber modes was studied, highlighting that the total channeled efficiency should take into account the efficiency of the QD itself [97]. To accurately measure this, they simultaneously recorded the photon count rates through both the guided and radiation modes of a nanofiber. QD sites were individually excited through an objective lens and the guided fluorescent photons were detected through the fiber pigtails. To factor in radiation mode guidance, fluorescent photons are also collected by the objective. Photon emission rates into both guided and radiation modes were calculated by analyzing photon-count rate histograms. By factoring in the light transmission parameters for both mode paths, the maximum channeling efficiency of (22.0 +/- 4.8)% obtained agreed with the theoretical predictions [47, 55]. It is expected that a photon channeling efficiency of higher than 90% could theoretically be obtained by incorporating a cavity structure on a nanofiber [96]. Other work related to couple of photons from nanoemitters into optical nanofibers is contained in Fujiwara et al. [98].

As QDs typically suffer from blinking and photobleaching, their potential as stable single photon sources is questionable. Nanodiamonds with nitrogen-vacancy-centers, which are free from these difficulties, have been deposited onto optical nanofibers by a dip-coating and translation technique. This scheme facilitates real-time estimation of the attached diamond numbers through fiber transmission monitoring [99]. Individual diamonds were illuminated through a microscope objective.

Red-shifted fluorescence spectra of these nanodiamonds were obtained through the fiber, and auto- and cross-correlation measurements showed anti-bunching of photons, again indicating that the light came from a single photon emitter. Combining nanofibers with unblinking color centers such as nanodiamonds could lead to fiber-coupled, on-demand single photon generation [100, 101]. The techniques described above also offer exciting possibilities for nanofiber detection and spectroscopy of biomolecules [102].

**Figure 10**. **(a)** Fluorescence counts depicted through a nanofiber as each QD deposition site is individually excited through a microscope objective; **(b)** Observed fluorescence count-rate for increasing intensities at positions 3,4,5 and 7 in (a), showing QD saturation. [96]

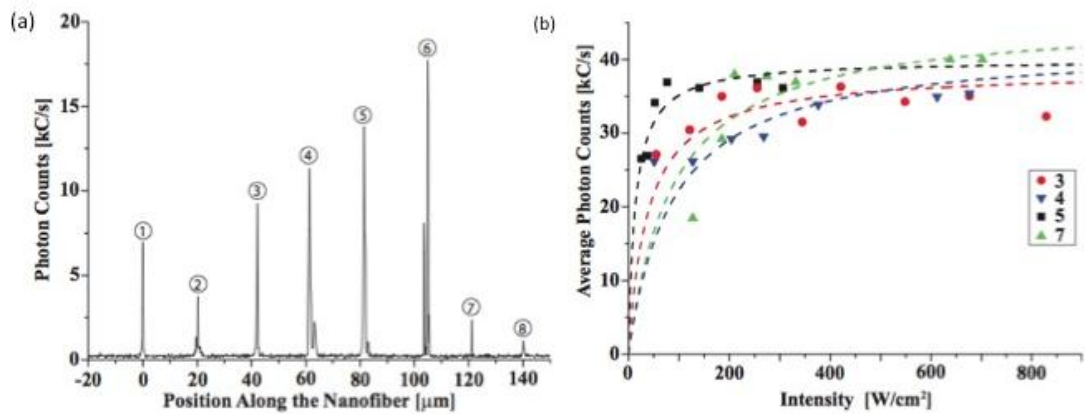

## 4. Alternative Techniques

The development of optical cavities has been of great interest in research for many years due to the significant enhancement effects and detection sensitivity offered by these devices [103]. With improved fabrication techniques various types of optical microcavities have been investigated [104-109]. Such microcavities have proven to be a productive platform for experiments with such effects as strong coupling [110] and the Purcell effect [111] being demonstrated. For detection purposes microcavities are extremely sensitive devices, arising from the low optical mode volume and long interaction times available due to the high-Q factors [103, 112], thereby enabling such devices to detect both single atoms [110] and single molecules [113].

Typically, for optical microcavities, the trend has been to use optical nanofibers to achieve coupling of light into and out of the cavity modes [110, 114-118]. Due to the advantages offered by cavities for detection a movement from these two element systems to a simpler system where the cavity is directly incorporated within the tapered optical fiber itself is underway. An example of this comes in the form of bottle resonators [16, 119-121], where the taper region itself is deformed to create a resonator, which generally still requires an external taper to couple light into the cavity. This shows how the tapered fiber can itself be used as a cavity. More direct examples of cavities within the fiber come in the form of fiber Bragg gratings (FBGs), for which much theoretical work has already been done on the interaction of a few/single atoms close to the nanofiber cavity [100, 122-126].

The fabrication of such FBGs has been developing in recent years from micron scale fibers down to nanometre scale fibers. For this, three fabrication techniques are used most frequently: (i) ultraviolet irradiation [127-130], (ii) focused ion beam (FIB) milling [101, 131-134], and (iii) femtosecond laser irradiation [135-138]. Each technique has advantages and disadvantages with quite possibly the most work to date concentrating on the FIB technique, see Fig. 11 for an example [101]. More recently, however, there has been success using femtosecond laser ablation to create highly ordered photonic crystal structures on a nanofiber, see Fig. 12 [138]. Such nanofiber cavities offer to be promising devices for the detection of very low numbers of atoms.

**Figure 11.** Scanning ion microscope image of a nanofiber Bragg grating fabricated using the FIB milling technique. The fiber diameter is ~560 nm. The grating period is ~360 nm, with each groove having a depth of ~100 nm and width of ~150 nm. [101] –

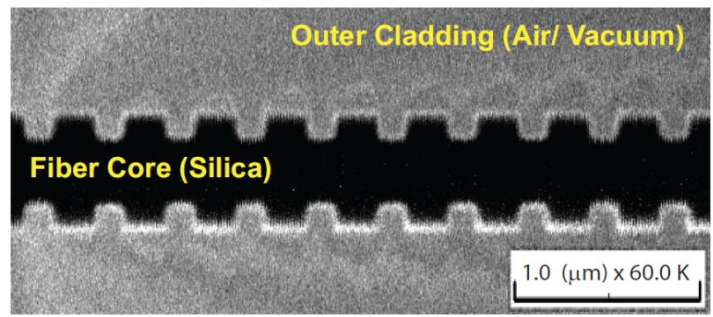

One problem with fixed cavities in the nanofiber region is tuning the cavity to useful wavelengths for detection purposes, which can be achieved, for example, by straining the fiber [139]. Other resonators which can be fabricated directly from tapered optical fibers and may avoid this problem are those where the fiber is looped upon itself. Examples of this are microcoils [140-142], microfiber knots [143, 144], and microfiber loops [145, 146]. With these cavities different tuning opportunities may be available [147-151]. For future possibilities using optical nanofibers, it has already been shown that polarization maintaining fiber can be tapered to micron sizes whilst still maintaining the polarization of the coupled light [152]. Such control and maintenance of polarization within the fiber may be of use when probing polarization sensitive states of atoms, or utilizing polarized confined modes within an optical cavity.

## 5. Conclusions/Outlook

The systematic fabrication of ONFs has now become readily available with several techniques already developed to achieve a smooth surface profile with low transmission losses and high mechanical strength making them ideal for the researcher to utilize in the development of quantum optical devices. In this article we have reviewed the state-of-the-art schemes that utilize ONFs for ultra-sensitive detection, trapping and manipulation of atoms and molecules. Due to the high coupling efficiency of atoms into the guided modes of the fiber, as well as the significant fraction of photons that can be absorbed from the evanescent field, ONFs provide an interface whereby the spectral properties of the atoms close to the surface of the fiber can be investigated. The ONF has the inherent ability to suppress ambient scattering of light, while maintaining the high coupling efficiencies of atoms. This allows the ONF to operate as a non-destructive detector to determine characteristics of atom clouds such as temperature, size and shape, as well as dynamic properties such as loading and decay times. Such techniques can have many future applications for non-destructive characterization of atomic ensembles. Based on the same principal, the ONF also acts as a sensor device for the detection of few atoms as well as single atoms with a high signal-to-noise ratio. These techniques have shown that ONFs have promise for numerous applications in quantum technologies where detection of small numbers of atoms will be an essential tool.

**Figure 12.** (a) Scanning electron microscope image of a sample fabricated using single-shot irradiation of a femtosecond laser beam. Inset: Enlarged view. Periodic nanocrater structures are observed on the shadow side of the nanofiber. (b) Cross-sectional image of a nanocrater measured by tilting the nanofiber. [138].

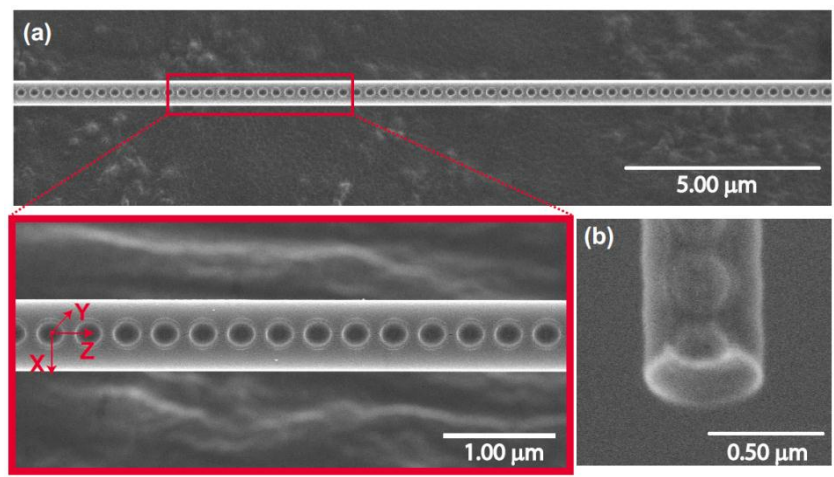

Due to the fact that atoms are detected close to the surface of the fiber, the detection schemes can facilitate fundamental research into physical phenomena such as the van der Waal and Casimir-Polder forces, which are otherwise more difficult to observe. These surface effects become important when considering the long-term performance or the degradation of the ONF in cold atom systems, whereby atoms become bound to the surface decreasing the performance of the ONF over time. This destructive effect can be reversed with the aid of a specific laser frequency passing through the fiber, thereby extending the lifetime of the ONF.

The unique properties of the ONF and its evanescent field allow various trapping schemes to be implemented. The configurations that have already been realized rely on two separate frequencies of light, with specific powers and polarizations to set up local optical lattice sites close to the surface of the ONF. The traps create strong confinement of single atoms in each lattice site. The atoms trapped in the sites can be translated by controlling the relative detuning of the standing waves, giving the ONF the ability to simultaneously, trap, probe and manipulate single atoms. This demonstrates the versatility and functionality of the ONF and paves the way for future ONF experiments such as coupling of atoms and entanglement via photon exchange.

Although the focus of this review has been on the utilization of ONFs with cold atoms, the methods and techniques can be easily extended to systems other than atoms. Molecules can be detected by performing surface absorption spectroscopy via the ONF. This has proven to be a more sensitive method when compared to free space techniques. In addition, for the use of an ONF as an interface with cold atoms, the structure of the ONF itself can be modified to incorporate devices such as grating and cavities. This micro- and nanostructuring of the ONF opens the path for multiple possibilities regarding the future of ONF as tools in the field of atom optics.

## 6. Conflict of Interest

"The authors declare no conflict of interest".